\documentclass[12pt]{article}
\usepackage{amsfonts,epsfig}
\usepackage{graphicx}
\usepackage{amsmath,amsfonts}
\usepackage{amssymb,graphicx}
\def\pa{\partial}
\def\L{\Lambda}
\def\s{\sigma}

\def\k{\kappa}

\def\d{\delta}
\def\r{\rho}
\def\h{\hat}

\def\dda{\ddot{a}}

\def\l{\label}
\def\e{\varepsilon}

\def\g{\gamma}
\def\m{\mu}
\def\n{\nu}
\def\be{\begin{equation}} 
\def\ee{\end{equation}}  
\def\ba{\begin{eqnarray}}
\def\ea{\end{eqnarray}}  

\def\h{\hat}
\def\b{\bar}
\def\t{\tilde}
\begin{document}  
\begin{flushright}
\end{flushright}
\vspace{10pt}
\begin{center}
  {\LARGE \bf Constraints on parameters of models with extra dimension 
from primordial nucleosynthesis } \\
\vspace{20pt}
M.Z.Iofa\\
\vspace{15pt}

\textit{Skobeltsyn Institute of Nuclear Physics, 
Moscow State University, Moscow,119992 , Russia}\\

    \end{center}
    \vspace{5pt}
\begin{abstract}
 5D models with one 3D brane and one infinite extra dimension are 
studied. 
Matter is confined to the brane, gravity extends to the bulk. 
 Models with positive and negative tension of the
brane are studied.
Cosmological solutions on the brane are obtained by solving the 
generalized Friedmann equation. As the input in cosmological solutions 
we use the present-time observational cosmological parameters.
We find constraints on
dimensionless combinations of scales of 5D models which 
follow from the 
requirement that with cosmological solutions we reproduce the 
data on production of ${}^4 He$  in primordial nucleosynthesis.
\end{abstract}
\section{Introduction}
Models with extra dimensions naturally appear in microscopic theories unifying 
gravity with other interactions. Viewed as models of the real world they 
must be tested as cosmological models. 
In this work we consider some aspects of cosmological solutions in 
five-dimensional with one 3D-brane embedded in the bulk
 with one infinite extra dimension \cite{maart}.
Matter is confined to the brane, and 5D gravity extends
to the bulk.

There are no strong constraints on scales of the models with extra
dimensions. 
 In the braneworld DGP models \cite{DGP} cosmology was discussed in a 
number of papers (the review article \cite{lue}, also 
\cite{DGKN2,DGKN3,def,DDG,DLRZA,lu.st}).  
In these studies the fundamental 5D 
scale varied in the range $ 10^{-12} GeV -  1 GeV $. The lower 
bound is 
provided by the tests of the Newton law, the upper bound follows from 
cosmology. 
Because the action of the DGP model does not contain  cosmological 
constants in the bulk and on the brane, the metrics obtained by solving 
the Einstein equations contain no warp factors.
 The metrics considered in these models were taken static and either 
asymptotically flat \cite{def,DDG,DLRZA},  or  deSitter \cite{lu.st}.

In this note we consider models which actions contain both bulk and 
brane cosmological constants. 
 In the Gaussian normal frame metrics in the bulk are warped 
solutions of the Einstein equations, satisfying  Israel junction
conditions on the brane with matter 
\cite{RS,BDL1,BDL2,Shtanov1,Shtanov2}.
The warp factor is taken in the range
$10^{-12} GeV - 10^{3} GeV$.

In models with extra dimension cosmological evolution on the brane is 
described by solutions of the non-standard Friedmann
equation \cite{BDL1,BDL2,Shtanov1,Shtanov2,def}.
We do not make a fit of parameters of models using the observational 
cosmological data, but
taking as the input the set of present-time observational cosmological 
parameters (Hubble
parameter, deceleration parameter and  fractions of cold matter and 
radiation in the total energy density) we look for constraints on 
parameters of the models which follow from the requirement that
the models yield abundance of ${}^4 He$ produced in
primordial nucleosynthesis within the experimental bounds.

We consider models with positive and negative tensions of the
brane.
The latter case is suggested by models with two branes, which allow for a
possibility to reconsider the hierarchy problem. In these models the visible
brane should have negative tension \cite{rub}.

Calculating production of ${}^4 He$ in the Big Bang nucleosynthesis 
(BBN)
we find constraints on dimensionless combinations of 
scales of the models. In particular, we 
find a relation
$$
\frac{\m M^2_{pl}}{M^3}\simeq \frac{2(1+q_0 )}{3\Omega_m}(\m
r_c \pm 1),
$$
where $\m$ and $M$ are the scale of the warp factor of the metric and 5D 
fundamental scale, and $r_c M^3$ 
is the coupling at the 4D Einstein term of the action.

\section{One-brane model}

We study one-brane models embedded in 5D bulk with the action
\be
\l{i1}
I_5  =\int\limits_\Sigma \,\sqrt{-g^{(5)}} \left(\frac{R^{(5)}}{2\k^2} +
\L\right) +\int\limits_{\pa {\Sigma}}
\,\sqrt{-g^{(4)}}\left(\frac{R^{(4)}}{2\k^2_1} -\h\s \right) -
\int\limits_{\pa
{\Sigma}} \sqrt{-g^{(4)}}L_m
,\ee
 where $\k^2 =8\pi/M^{3}$, $\k^2_1 =8\pi/ r_c M^3$.
$L_m$ is the Lagrangian of matter on the brane.    
$M$ is the gravitational scale of the 5D gravity, parameter $r_c$ defines
the strength of gravity in the 4D term. 

It is assumed that 
matter is confined to the spatially flat brane and  gravity is 5-dimensional. 
To have homogeneous cosmology on the brane the energy-momentum 
tensor of matter on the brane is taken 
in a phenomenological form
\be
\l{2}
 T^\m_\n =diag\{-\h{\r},\h{p},\h{p},\h{p}\},
\ee
where $\h{\r}(t)$ and $\h{p}(t)$ are the sums of energy densities and
pressures of cold matter and radiation.
For the following it is convenient to introduce
the normalized expressions for bulk cosmological constant, energy
density, pressure and
cosmological constant on the brane which all have the same
dimensionality $GeV$
\be  
\l{4}
\m =\sqrt{\frac{\k^2\Lambda}{6}},
\quad \r =\frac{\k^2\h\r}{6},    
\quad \s =\frac{\k^2\h\s}{6},\quad {p} =\frac{\k^2\h{p}}{6}.
\ee
The Einstein equations are solved in a class of metrics of the form
\be
\l{3}
 ds^2_5 = -n^2 (y,t)dt^2 + a^2 (y,t)\g_{ij}dx^i dx^j +dy^2 .
\ee
The brane is located at the fixed position $y=0$. Using the freedom in
parametrization of time, the component of the metric $n(y,t)$ is normalized
so that $n(0,t)=1$. 

We study cosmologies on the brane by solving the non-standard Friedmann
equation \cite{BDL1,BDL2,Shtanov2,def}
 which follows from the system of the Einstein equations and junction
(Israel) conditions on the brane (in the following we assume the symmetry
$y\rightarrow -y$)
\ba  
\l{5}
&{}& \frac{a'(0,t)}{a(0,t)}= -\s -\sum\r_i (t)
+\frac{r_c}{2}\frac{\dot{a}^2(0,t)}{a^2(0,t)},
\\\nonumber
&{}& \frac{n'(0,t)}{n(0,t)} =-\s +\sum (2\r_i +3p_i ) +\frac{r_c}{2}
\left(-\frac{\dot{a}^2(0,t)}{a^2(0,t)}
+2\frac{\ddot{a}(0,t)}{a(0,t)}\right), 
\ea
 where $\r_i , p_i \,\, i=r,m $ are densities and pressures of the
cold matter and radiation.
Equation for the scale factor $a(0,t)$ can be obtained either in the 
form with the second-order
derivatives of the scale factor \cite{BDL1},
\be  
\l{6}
\frac{\ddot{a}(0,t)}{a(0,t)} +\frac{\dot{a}^2 (0,t)}{a^2 (0,t)} = -2\mu^2 -
\left(\s + \sum\r_{i} -\frac{r_c}{2}\frac{\dot{a}^2 (0,t)}{a^2 (0,t)}\right)  
\left(-2\s + \sum\r_{i} +3 \sum p_{i} + r_c\frac{\ddot{a}(0,t)}{a(0,t)}
\right)
,\ee
 or, starting from the partially
integrated system of the Einstein equations, in the form with the
first-order derivatives \cite{BDL2,Shtanov1,Shtanov2,def}
\be
\l{8}
\frac{\dot{a}^2 (0,t)}{a^2 (0,t)}=-\m^2 +
\left(\s + \sum\r_{i} -\frac{r_c}{2}\frac{\dot{a}^2 (0,t)}{a^2
(0,t)}\right)^2 + C\frac{a^4 (0, t_0 )}{a^4 (0, t)}
,\ee
where the last term is interpreted as dark radiation 
\cite{maart,maeda}.
Introducing notations
$$
z =a(0, t_0 )/a(0, t)-1,\quad H(t)=\dot{a} (0,t)/a (0,t), \quad
C(z+1)^4\equiv\m\r_w (z),
$$ 
where $t_0$ is the present time,
Eq.(\ref{8}) can be written as 
\be
\l{8a}
(1+\s r_c )H^2 = \s^2-\m^2 +2\s (\r_m +\r_r ) + \left(\r_m +\r_r -\frac{r_c
H^2}{2}\right)^2 + \m\r_w 
\ee
Energy densities of cold matter and radiation are
\be  
\l{11}
{\r}_m ={\r}_{m0} (1+z)^3, \quad {\r}_r ={\r}_{r0}r(z) (1+z)^4 ,
.\ee
where $r(z)$ is a slow function of $z$ which counts the number of
relativistic degrees of freedom, $r(0)=1$.
If expressed as a function of temperature of the Universe, $\hat{\r}_r 
(T) =\pi^2/30 g_* (T)T^4$ \cite{kolb}. 
This makes possible to connect $r(z)$ and $g_* (T)$.
\subsection{Model with positive tension of the brane}

We define the period of   "late cosmology"  as 
a period in which in the Friedmann
Eq.(\ref{8a})
the terms linear in matter (radiation) energy density are dominant
\be
\l{lc1}
\s\r(z) > \r^2 (z), \quad \s\r(z) > r_c H^2 \r(z), \quad \s\r(z) > (r_c
H^2 )^2 .
\ee
Friedmann Eq. (\ref{8a}) can be approximately written as
\be
\l{lca}
(1+r_c \s )H^2 \simeq \s^2 -\m^2 +2\s (\r_m (z) +\r_r (z) ) +\m\r_w (z)
.\ee
Using (\ref{lca}) it can be shown that the second and the third conditions 
(\ref{lc1}) follow 
from the first condition $\s >\r (z)$.

From Eqs. (\ref{6}) and (\ref{8}) taken at present time we express $\s^2
-\m^2$ and $C$. Introducing dimensionless combination
\be
\l{dp1}
p^2 =\frac{H^2_0}{\s \r_{m0}}
\ee 
and neglecting
the terms of the second order in the present-time energy densities, we 
have 
\be
\l{n4a}
\s^2-\m^2 \simeq \frac{H^2_0}{2p^2}\left[(1+r_c\s  )(1-q_0 )p^2 -1 \right]
,\ee    
and     
\be     
\l{n8}  
C\simeq 
\frac{H^2_0}{2p^2}\left[-3-4\frac{\Omega_r}{\Omega_m} +
(1+r_c\s )(1+q_0 )p^2 \right],
\ee
where  $q_0 =-\dda(t_0 )/a(t_0 )H_0^2$ is the present-time deceleration 
parameter.

Substituting in (\ref{lca}) expressions (\ref{n4a}) and (\ref{n8}), 
we obtain Friedmann equation in the period of late 
cosmology
\ba   
\l{n9}
&{}&H^2\simeq \frac{H^2_0}{2}\left[1-q_0 +(1+q_0 )(z+1)^4 
\right.\\\nonumber
&{}&\left.+\frac{1}
{p^2 (1+\s r_c )}\left(-1+4(z+1)^3 -
\left(3+4\frac{\Omega_r}{\Omega_m} (r(z)-1)\right)(z+1)^4 
\right)\right]
.\ea
The rhs of (\ref{n9}) is positive if
\be    
\l{n10}
p^2 (1+ \s r_c  )(1+q_0 )>3
.\ee
We assume that parameter $r_c$ is constrained so that 
\be
\l{lc3}
  \m r_c  \frac{H^2_0}{\m^2}\ll 1
.\ee
For the smallest  $\m\sim 10^{-12} GeV$  we obtain the estimate 
$\m r_c <10^{60}$.
Below it will also be shown that $p^2 (1+ \s r_c  )(1+q_0 )-3\ll 1$, and 
hence 
\be
\l{dp2}
p^2\simeq \frac{3}{(1+\s r_c )(1+q_0 )}
.\ee 
Using the constraints (\ref{n10}) and (\ref{lc3})  in equation
 (\ref{n4a}), we obtain that
\be
\l{lc4}
\s =\m +O\left(\m r_c \frac{ H^2_0} {\m}\right)
.\ee
 With this accuracy we substitute $\m$ for $\s$.
At large $z$ the first condition of late cosmology (\ref{lc1}) is
$$
\m>\r_{r0}r(z) z^4 =\r_{m0}\frac{\Omega_r}{\Omega_m}r(z)z^4 .
$$
Using (\ref{dp1}) and (\ref{dp2}) this condition  can be 
written as
\be
\l{lc8}
z^4 r(z) \ll \frac{\m^2 p^2 \Omega_m }{\Omega_r H^2_0}\sim \frac{\m^2}{\Omega_r 
H^2_0 (1+\m r_c )}.
\ee
In the model with $r_c =0$, in the radiation-dominated period, Friedmann 
equation is
\be
\l{l15}
H^2 =\frac{H^2_0}{2p^2} \left(-3 +4\frac{\Omega_r}{\Omega_m}(r(z)-1)+
(1+q_0 )p^2 \right) z^4 +\left( 
\frac{H^2_0}{p^2 
\m}\frac{\Omega_r}{\Omega_m}r(z)z^4 
\right)^2
,\ee 
where we have kept both the linear and quadratic terms.
The linear term is dominant, if
$$
1\gg \frac{H^2_0}{\m^2}\Omega_r r^2 (z) z^4,
$$
which is the same as condition of late cosmology (\ref{lc8}) 
\be
\l{l16}
\Omega_r^{-1}< z< \left(\frac{\m^2}{H^2_0 \Omega_r r(z)}\right)^{1/4}.
\ee

Because the present-time total energy density of the Universe is 
approximately equal to the critical energy density $\h{\r}_c =3H^2_0 
M^2_{pl}/8\pi$, we can express $p^2$ as 
$$
p^2 \simeq\frac{H^2_0}{\m\Omega_m\r_c}=\frac{2M^3}{\Omega_m\m M^2_{pl}}
,$$
where $\r_c =\k^2 \h{\r}_c/6$.
From the relation $p^2 (\m r_c +1) \simeq 3/(1+q_0 )$  we obtain
\be
\l{n11a}
\frac{\m M^2_{pl}}{M^3} \simeq \frac{2(1+q_0 )}{3\Omega_m}(\m
r_c +1).
\ee
Here the combination $ 3\Omega_m/2(1+q_0 ) $ is of order unity
\footnote{
With $q_0 \simeq -1 +3\Omega_m /2 $ \cite{part,harun,steig}, one obtains
$3\Omega_m/2(1+q_0 )\simeq 1$.}.

\section{Model with negative tension of the brane}
In the model with negative tension of the brane we set in (\ref{6}) and 
(\ref{8})  $\s =-|\s|$. The period of late cosmology is defined by 
conditions
$$
|\s|\r(z) > \r^2 (z), \quad |\s|\r(z) > r_c H^2 \r(z), \quad |\s|\r(z) > 
(r_c H^2 )^2 
$$
ensuring that in the Friedmann equation the terms linear in energy 
densities are dominant. 
Taking  Eqs. (\ref{6}) and (\ref{8}) at present time 
and  omitting the terms of the second order in energy densities, we have
\ba 
\l{n2b}
H^2_0 (1- r_c |\s| )(1-q_0) \simeq 2(\s^2 - \m^2 ) -|\s| {\r}_{m0}, \\
\l{n3b}
H^2_0 (1- r_c |\s| ) \simeq
-\m^2 + \s^2  -2|\s| ({\r}_{m0} +\r_{r0} )  +C
.\ea 
From these relations we express $\s^2 -\m^2$ and $C$
\ba
\l{n4b}
\s^2 -\m^2 \simeq\frac{H^2_0}{ 2p^2}\left(1+(1-q_0 )(1-|\s| r_c  )p^2\right),\\
C\simeq\frac{H^2_0}{2p^2}\left[3+4\frac{\Omega_r}{\Omega_m} +
(1-|\s| r_c  )(1+q_0 )p^2 \right]
.\ea
In the model with $r_c =0$, in the radiation-dominated period, keeping
both the terms linear and quadratic in the radiation energy density
we obtain the Friedmann equation as
\be
\l{nb7}
H^2 \simeq 
\frac{H^2_0}{2p^2}\left(3+(1+q_0 
)p^2 \right) z^4 +
\left(\frac{H^2_0}{p^2\s}\right)^2
\left(z^3 +\frac{\Omega_r}{\Omega_m}r(z) z^4\right)^2
,\ee
where we neglected the term $(r(z)-1)\Omega_r/\Omega_m$ as compared to 
$3+ (1+q_0 )p^2$.

In the model with $r_c\neq 0$, neglecting quadratic terms and setting 
$r(z)=1$, the 
Friedmann equation in the period of late cosmology is
\be
\l{5a}
H^2\simeq \frac{H^2_0}{2}\left[(1-q_0 ) +(1+q_0 )(1+z)^4
+\frac{1}{{p}^2 (1-|\s| r_c )}\left(1-4(z+1)^3 +3(z+1)^4 \right)\right]
.\ee
In the case $1-|\s| r_c >0$,
for $|q_0 | <1$
the rhs of (\ref{5a}) is a positive increasing function
for all $z >0$, and there are no
constraints on $p^2$.

In the case $|\s| r_c -1 >0$ the
 Friedmann equation
 in the period of late cosmology can be written as
\be
\l{5b}
H^2\simeq \frac{H^2_0}{2}\left[(1-q_0 ) +(1+q_0 )(1+z)^4
+\frac{1}{{p}^2 (|\s| r_c -1)}\left(-1+4(z+1)^3 -3(z+1)^4 \right)\right]
.\ee
Equation (\ref{5b}) has the same functional form as that in the model
with positive  tension.
The rhs of (\ref{5b}) is positive for
\be    
\l{n7b}
p^2 (\m r_c -1 )(1+q_0 )>3
.\ee
Constraining $\m r_c$ so that $1\gg (\m r_c )H^2_0 /\m^2$, we can show 
that
$$
|\s| =\m +O\left(\m r_c \frac{H^2_0}{\m}\right).
$$

\section{Primordial nucleosynthesis}
\subsection{Model with positive tension of the brane}
In this section, comparing predictions of BBN in the standard and
non-standard cosmologies, we obtain constraints on parameters of the
non-standard model. 
First, we study the model with positive
tension of the brane without the $4D$ curvature term in the action.

In the standard cosmology, in the
radiation-dominated period,  the radiation energy density expressed as a
function of $t$ and of temperature of the Universe $T$ is
\ba
\l{p1}
&{}&\h{\r}_r  (t) =\h{\r}_{r0}r(z)z^4
=\frac{\h{\r}_{r0}}{4\Omega_r (H_0 t)^2},\\\nonumber
&{}&\h{\r}_r  (T)=\frac{\pi^2}{30}g_{*}(T) T^4.
\ea
In the non-standard model, in the period of late cosmology we have
\ba
\l{p3}
&{}&\t{\h{\r}}_r  (t)\simeq\frac{\h{\r}_{r0}r(z)}{2(H_0 t)^2}
\frac{p^2}{[(1+q_0 )p^2 -3 +4(r(z)-1)\Omega_r/\Omega_m ]},\\\nonumber
&{}&\t{\h{\r}}_r (\t{T})=\frac{\pi^2}{30}g_{*}(\t{T}) \t{T}^4
\ea
where by {\it tilde} we  distinguish
  the non-standard case. Here $z=z(t)$. 

The freezing temperature $T_F$
of the reaction $n\leftrightarrow p$ is estimated as a temperature at which
the Hubble parameter $H$ is of order of the reaction rate $G_F^2
T_F^5$ \cite{kolb}. Time dependencies of the Hubble parameter in both the standard
and
non-standard cosmologies are the same $H=1/2t$. 
Using (\ref{p1}) and (\ref{p3}), we obtain the ratio of freezing
temperatures in the standard and non-standard cosmologies
\be   
\l{p4}
\frac{\t{T}_F}{T_F}=\left(\frac{{g}_{*}(\t{T}_F) }{{g}_{*}(T_F )}
\frac{[(1+q_0 )p^2 -3 +4(r(z)-1)\Omega_r/\Omega_m ]}
{2p^2 r(z)\Omega_r }\right)^{1/6}
.\ee
The mass fraction of  ${}^4 He$ produced in nucleosynthesis of the total
baryon mass is
\be   
\l{p51}
 X_4 = \frac{2(n/p)_f}{(n/p)_f +1} 
,\ee
where the subscript "f" indicates that the ratio is taken at the end of
primordial nucleosynthesis, $(n/p)_f \simeq 1/7$ \cite{kolb}.
The equilibrium value of the
neutron-proton ratio $(n/p)_{T_F}=\exp{[-(m_n -m_p )/T_F]}$ is very
sensitive
to the value of $T_F$. 
Under variation of freezing temperature variation of $X_4$ is
\be
\l{p6}
\delta X_4\simeq \frac{2}{((n/p)_f +1)^2} ({n}/{p})_f\ln
({p}/{n})_F \frac{\d T_F}{T_F}.
\ee
We constrain parameters of the model requiring that the difference between
the calculated values of $X_4$ in the standard and non-standard models 
is within the experimental errors. 
The estimate of $X_4$ obtained in the standard cosmology $X_4 \simeq 
0.25$ fits
well the experimental value $X_4 =0.25\pm 0.01$ \cite{part}.
Using (\ref{p6}) we estimate the variation of the freezing temperature 
retaining $X_4$ within the experimental errors
$$
\frac{\d T_F}{T_F}< 0.0255.
$$
Comparing $T_F$ and $\t{T}_F$, we have
$$
\frac{\d T_F}{T_F} \simeq\left( \frac{{g}_{*}(\t{T}_F) }{{g}_{*}(T_F )}
\frac{[(1+q_0 )p^2 -3 +4(r(z)-1)\Omega_r/\Omega_m ]}
{2p^2 r(z)\Omega_r }\right)^{1/6} -1.
$$
From this relation follows an estimate
\be
\l{p7}
\left|\frac{{g}_{*}(\t{T}_F) }{{g}_{*}(T_F )}
\frac{[(1+q_0 )p^2 -3 +4(r(z)-1)\Omega_r/\Omega_m ]}
{2p^2 r(z)\Omega_r }-1 \right| <\e 
\ee
where $\e=0.164$.
It follows that
\be
\l{p7a}
(1+q_0 )p^2 -3= O(\Omega_r ).
\ee
Introducing dimensionless ratio of scales
$\m M^2_{pl}/M^3 $,
we can write
$$
 p^2\simeq \frac{2M^3}{\Omega_m \m M^2_{pl}} .
$$
Substituting this relation in (\ref{p7}), we obtain an estimate
\be
\l{p9}
\left(1-\e -\frac{2(1+q_0 )(r-1)}{3\Omega_m r}\right)
<\frac{\m M^2_{pl}}{M^3} -\frac{2(1+q_0 )}{3\Omega_m}<
\frac{4r\Omega_r}{3\Omega_m}\left(1+\e -\frac{2(1+q_0 )(r-1)}
{3\Omega_m r}\right). 
\ee
 
Now we can verify that in the non-standard model BBN takes place at the
period of late cosmology.
From condition of late 
cosmology  (\ref{lc8}),  the transition 
point to late cosmology, can be bounded as
\be
\l{z}
\bar{z}^4 <\frac{\m^2}{H^2_0\Omega_r}.
\ee 
In the most stringent case, for $\m\sim 10^{-12} GeV$, this yields $\bar{z}\sim 
10^{16}$ which is much larger than $z_{BBN}\sim 10^{9\div 10}$.
In the radiation-dominated period, setting $r(z) =1$, we obtain solution 
of the Friedmann equation as 
\be
\l{p8}
z^{-4}\simeq\frac{2[(1+q_0 ){p}^2 -3 ]}{{p}^2} (H_0 t)^2 +
\frac{4 \Omega_{r}}{{p}^2\,\mu \Omega_{m}}H^2_0 t.
\ee
From (\ref{p8}) and (\ref{z}) it follows that the transition time 
to late cosmology is $\b{t}\sim 1/\m \sim 10^{12} GeV^{-1}$. 
This is much smaller than  characteristic time of nucleosynthesis
 $1\div 10^2 s$, or $10^{24\div 26} GeV^{-1}$.

Let us turn to the model with 4D curvature term included in the action.
Following the same steps as in the model with $r_c =0$, we find that
production of Helium is within the experimental bounds 
if
\be
\l{p9b}
\left|\frac{p^2 (\m r_c +1 )(1+q_0 )-3+4(r-1)\Omega_r/\Omega_m }{2p^2 
(\m r_c +1)r\Omega_r} -1\right|< \e
,\ee
where $\e \simeq 0.164$.
From (\ref{p9b}) follows the constraint on $p^2$
\be
\l{p9c}
\frac{4r\Omega_r}{\Omega_m}
\left((1-\e)\frac{3\Omega_m}{2(1+q_0 )} -\frac{r-1}{r}\right)
<p^2 (1+\m r_c )(1+q_0) -3 <\frac{4r\Omega_r}{\Omega_m}\left((1+\e 
)\frac{3\Omega_m}{2(1+q_0 )} -\frac{r-1}{r}\right) 
,\ee
which can be written as
\be
\l{p9e}
p^2 (1+\m r_c )= \frac{3}{1+q_0 }\left(1 +O(\Omega_r ) \right)
.\ee
Using the inequalities (\ref{p9c}) we obtain the bounds on the 
(normalized) density of dark radiation $\r_{w0}$ 
\be
\l{p11}
\frac{H^2_0}{\m^2} \Omega_r (1+\m r_c )
\left(1-\e- \frac{2(1+q_0 )}{3\Omega_m}\right)
<\frac{\r_{w0}}{\m}<\frac{H^2_0}{\m^2} 
\Omega_r (1+\m r_c )r\left(1+\e- \frac{2(1+q_0 )}{3\Omega_m}\right).
\ee
The combination $2(1+q_0 )/3\Omega_m$ is of order unity, and within the
existing uncertainties of cosmological parameters the sign of $\r_{w0}$ is
ambiguous. From (\ref{p11}) it follows that  $\r_{r0}/\m > \r_{w0}/\m$.
At times at the beginning of nucleosythesis the effective number of 
degrees of freedom is $g_* 
(T_{BBN})=10.75$, and we obtain $r\simeq 3.2$.

Solving the
Friedmann equation (\ref{n9}) with $r(z)=1$ in the radiation-dominated 
period of late cosmology, we have
\be
\l{p10}
z^{-4}\simeq\frac{2[p^2 (\m r_c +1 )(1+q_0 )-3 ]}{p^2 (\m r_c +1)}
(H_0 t)^2 .
\ee
In (\ref{lc8}) the period of late cosmology was defined as 
$$
z^4 \ll \bar{z}^4 \sim \frac{\m^2}{\Omega_r H^2_0 (1+\m r_c )}.
$$
For the characteristic times of nucleosynthesis, $t_{BBN}\sim 10^{24\div 
26} GeV^{-1}$, we have $z_{BBN}\sim 10^{9\div 10}$.
Requiring that $z_{BBN}\ll \bar{z}$, we obtain
\be
\l{p16}
\frac{H^2_0}{\m^2}\m r_c \ll \frac{1}{\Omega_r z_{BBN}^4}
.\ee
In the most stringent case, for $\m\sim 10^{-12} GeV$, from (\ref{p16}) 
it follows that
$$
\m r_c \ll 10^{25},
$$
which is stronger than the bound (\ref{lc3}).
\subsection{Model with negative tension of the brane}
First, we consider the model without the 4D curvature tern in the action.
 In the radiation-dominated period, solution
of Friedmann equation is
\be
\l{p13}
z^{-4}\simeq\frac{2(3+(1+q_0 ){p}^2 )}{p^2} (H_0 \t{t})^2 +
\frac{4 \Omega_r}{p^2\,\mu \Omega_m}H^2_0 \t{t},
\ee
where we neglected the slow factor $r(z)$. 
At the characteristic times of nucleosynthesis
 the first term in (\ref{p13}) is dominant, and we can neglect the
second one.

Following the same steps as in the model with positive tension of the brane,
we obtain the ratio of the freezing temperatures  in the
standard and non-standard cosmologies    
\be   
\l{p14a}
\frac{\t{T}_F}{T_F}=\left(\frac{{g}_{*}(\t{T}_F) }{{g}_{*}(T_F )}
\frac{[3+(1+q_0 )p^2]}{2p^2\Omega_r }\right)^{1/6}
.\ee
Because $(3+(1+q_0 )p^2 )/2p^2\Omega_r $ is a large number, 
the ratio $\t{T}_F/T_F$ is not close to unity, and  
the abundance of $He$ calculated in this  model is
 outside the experimental bounds. 
With the minimal value of the factor 
$[3+(1+q_0 )p^2]/2p^2\Omega_r\sim 10^4$ we obtain $\t{T}_F/T_F =4.64$ and 
$(n/p)_{\t{T}_F}=0.68$. 

Using (\ref{p13}) and (\ref{p1})  we find that the times $t$ and $\t{t}$ 
 the Universe has the same temperature $T$ both in 
the standard and non-standard models are 
connected as
$$
\t{t} =t\left(\frac{\Omega_r p^2}{p^2(1+q_0 )+3}\right)^{1/2} 
.$$
It follows that $\t{t}\lesssim 10^{-2} t$. In the non-standard model the 
interval of 
times  
at which the Universe is cooled from
the temperature $\t{T}\sim 1 MeV$ to  $0.1 MeV$  between 
the beginning and the end of the BBN is $\sim 1 s$. The decay of neutrons 
during this time interval is negligible as compared with the standard 
model.
The fraction $X_4$ can be calculated at the freezing point
\be
\l{p15}
X_4 = \frac{2(n/p)_F}{(n/p)_F +1} \simeq 0.81
,\ee
and is considerably larger than the observational value.

In the model with 4D curvature term included in the 5D action we considered
two cases. If $1-|\s| r_c >0$, the model is similar to the case without 
4D
term. As it was discussed above, this case the model 
yields the value of $X_4$ mach larger than the observational bound.
If $|\s | r_c -1 >0$, solutions of the Friedmann 
equations are similar to those in the
model with positive tension of the brane and in this case the value of 
$X_4$ is within the experimental bounds if parameters of the model are 
restricted as
$$
\left|\frac{p^2 (\m r_c -1 )(1+q_0 )-3+4(r-1)\Omega_r/\Omega_m }{2p^2
(\m r_c -1)r\Omega_r} -1\right|< \e
.$$

\section{Conclusions}
In this note we discussed a class of warped cosmological solutions of 
the Friedmann equation in 
one-brane 5D models with infinite extra dimension.
We considered models with positive and negative tensions of the
brane. 
We studied separately the  models  with and without 4D
curvature term on the brane included in the 5D action.   
We found constraints on parameters of the
models which follow from consistency of predictions of the models on 
abundance of ${}^4 He$ produced  in primordial
nucleosynthesis with  the cosmological data.

For numerical estimates we used the  present-time values of 
the Hubble parameter $H_0$, fractions of cold matter and radiation
$ \Omega_m , \Omega_r$,
deceleration parameter $q_0$, and the mass fraction $X_4$ of ${}^4 He$ 
of 
 the total barion mass produced  in primordial
nucleosynthesis.

It was found that in the models with positive tension of the brane, both 
with 
and without 4D curvature term, production of ${}^4 He$ can be consistent 
with the cosmological data, if parameters of the models satisfy the 
constraint
$$
\left|\frac{\m M^2_{pl}}{M^3 (\m r_c +1)} -\frac{2(1+q_0
)}{3\Omega_m}\right|< O(\Omega_r ).
$$ 
Because $3\Omega_m /2(1+q_0 )\sim 1$, it follows that 
$$
 \frac{\m M^2_{pl}}{M^3}\sim {\m r_c +1}
,$$
where an upper bound on $\m r_c$ is 
$$
\m r_c  < 10^{25}
.$$
In the models with negative tension of the brane the results depend on 
whether 4D curvature term is included in the action of the 5D model or 
not.
The model  without 4D  curvature term in the action
 yield abundance of ${}^4 He$ considerably higher than  the 
observational data.

The model with 4D curvature term
included in the 5D action can meet the observational data provided
$\m r_c >1$.
From the estimate of the abundance of ${}^4 He$ follows a constraint 
on parameters
\be
\l{con}
\left|\frac{\m M^2_{pl}}{M^3 (\m r_c -1)} - 
\frac{2(1+q_0 )}{3\Omega_m}\right|< C\Omega_r,
\ee
where $C\sim 10$.

Further restrictions on parameters of the model with negative tension of 
the brane can be obtained, if  
 $M_{pl}^2$ is larger than the coupling at the 4D term in the action, 
i.e. if $M^2_{pl}>r_c M^3$, or $\m M^2_{pl}/M^3>\m r_c$.
In this case
$$
\m r_c < \m M^2_{pl}/M^3 <(\m r_c -1) 
\left(\frac{2(1+q_0 )}{3\Omega_m}  +C\Omega_r \right)
,$$
or  
$$
\frac{2(1+q_0 )}{3\Omega_m}<\m r_c\left( \frac{2(1+q_0 
)}{3\Omega_m}+ C\Omega_r -1 \right).
$$
Because $2(1+q_0 )/3\Omega_m\sim 1$, it follows that $\m r_c \gg 1$ and 
hence $\m M^2_{pl}/M^3\gg 1$.

Generally, numerical values of cosmological parameters 
depend on a model in which 
 the experimental data were processed. Recently fits of 
parameters were
made in RSII-like models with positive tension of the brane
\cite{dab,hung1,hung2,fay}. 
The main cosmological parameters obtained in these fits do 
not considerably differ from those of the standard model used in the 
present note. With these 
parameters qualitative 
conclusions of the present paper remain valid.

\end{document}